\documentclass[a4paper]{PoS}

\title{VERITAS Observations of HESS J1943+213}

\ShortTitle{VERITAS Observations of HESS J1943+213}

\author{\speaker{Karlen Shahinyan} for the VERITAS Collaboration\thanks{veritas.sao.arizona.edu}\\
        University of Minnesota\\
        E-mail: \email{shahin@astro.umn.edu}}

\abstract{HESS J1943+213 is a very-high-energy (VHE; \textgreater 100 GeV) gamma-ray point source detected during the H.E.S.S. Galactic Plane Survey. Radio, infrared, X-ray, and GeV gamma-ray counterparts have been identified for HESS J1943+213; however, the classification of the source is still uncertain. Recent publications have argued primarily in favor of an extreme BL Lac object behind the Galactic plane, though the scenario that HESS J1943+213 is a young pulsar wind nebula is viable as well. We present deep VERITAS observations of HESS J1943+213, which provide the most significant VHE detection of the source so far, with $\sim$18 sigma excess. The source is detected at $\sim$2\% Crab Nebula flux above 200 GeV with VERITAS, with the source spectrum well fit by a power-law function and showing agreement with the H.E.S.S. detection. We also include results from analysis of \emph{Swift} XRT observations contemporaneous with VERITAS. No significant flux or spectral variability is detected with VERITAS or \emph{Swift} XRT observations. We place the VERITAS results in a multi-wavelength context to comment on the HESS J1943+213 classification.}

\FullConference{The 34th International Cosmic Ray Conference,\\
		30 July- 6 August, 2015\\
		The Hague, The Netherlands}

\begin{document}

\section{HESS J1943+213 Classification}
HESS J1943+213 was discovered in very-high-energy (VHE; E $\textgreater$ 100~GeV) gamma rays during 
the H.E.S.S. Galactic plane scan~\cite{hess}. Due to its point-like appearance in VHE gamma rays, three possible
source classes were suggested: gamma-ray binary, pulsar wind nebula (PWN), and BL Lac object (blazar). 

Assuming the source is a gamma-ray binary, ref.~\cite{hess} used the lack of detection of a massive (O- or B-type) companion star to estimate a distance limit of greater than $\sim$25 kpc. This distance would place the binary well beyond the extent of the Galactic disk and would imply an X-ray luminosity 100-1000 times higher than luminosities of known gamma-ray binaries. Such a distance limit is problematic and this scenario is disfavored. In addition, the point-like appearance in the X-rays and the soft VHE spectrum with a power-law index of $\Gamma$=3.1 $\pm$ 0.3 (in contrast to the softest known PWN index of 2.7) motivated ref.~\cite{hess} to argue against the PWN scenario, leaving the blazar hypothesis. The authors found all observations to be consistent with the blazar scenario: the point-like nature in both X-rays and VHE, the soft VHE spectral index, and a preliminary IR spectrum showing lack of emission lines. Moreover, the hard X-rays observed with INTEGRAL IBIS and \emph{Swift} BAT instruments showed no evidence of a cutoff up to an energy of $\sim$195~keV. If the source is a blazar, it would be categorized as an extreme high-synchrotron-peak BL Lac object (extreme HBL), a class of blazars with the synchrotron peak located at energies \textgreater1 keV~\cite{ehbl}.

Since the discovery publication, the identity of HESS J1943+213 has been the topic of an ongoing debate. 1.6-GHz VLBI observations of the HESS J1943+213 counterpart with the European VLBI Network produced
a detection that was claimed to show extension, with FWHM angular size of 15.7~mas (with 3.5~mas expected size for a point source)~\cite{gabanyi}. 
 Based on this measurement, the brightness temperature of the counterpart was estimated to be 7.7$\times$10$^{7}$ K and was used to argue against the blazar scenario, as the expected brightness temperature of HBLs is in the 10$^{8}$--10$^{9}$ K range. In addition, ref.~\cite{gabanyi} employed a 1$'$ feature observed in the 1.4-GHz VLA C-array configuration image to support the PWN hypothesis, with the assertion that the angular size of the feature is consistent with a Crab-like PWN placed at a distance of 17~kpc. On the other hand, ref.~\cite{tanaka} argued in favor of an extreme HBL by constructing a spectral energy distribution and drawing comparisons to a known extreme HBL, 1ES 0347-121. More recently, the case for the extreme blazar scenario has been bolstered with ref.~\cite{peter} observing the near-infrared (K-band) counterpart of HESS J1943+213, and claiming potential detection of an elliptical host galaxy with only 10\% probability the object is a star. Using 5 years of data, they also obtain the first significant detection of the source with Fermi LAT (within 2$\sigma$ positional uncertainty) in the 1--300 GeV energy regime and use it in conjunction with the H.E.S.S. spectrum to set an upper limit on the redshift at z < 0.4. 
Finally, ref.~\cite{gab2015} obtained follow-up observations in the 1.5 GHz and 5 GHz bands using the e-Multiple Element Remotely Linked Interferometer Network (e-MERLIN), showing that the source exhibits a flat spectrum between the two bands and claiming a detection of flux density variability in the 1.5 GHz band when compared with the EVN observations of the source. 
Though recent evidence is stacking in favor of the blazar hypothesis, there is yet to be a definitive identification of the HESS J1943+213 source class.

Here we present results from VERITAS observations of HESS J1943+213 and discuss implications for the debate regarding the source identity.

\section{VERITAS Observations}
The Very Energetic Radiation Telescope Array System (VERITAS) is an array of four 12-m imaging 
atmospheric Cherenkov telescopes, located in Arizona, USA at an elevation of $\sim$1270~m. The camera for 
each telescope is composed from 499 photo-multiplier tubes, with a field of view of approximately 3.5$^{\circ}$.
VERITAS is able to reliably reconstruct VHE gamma rays with energies between 85~GeV and \textgreater30~TeV~\cite{holder}, with an energy resolution of 15-25\% and a sensitivity to detect a 1\% Crab Nebula flux source at 5$\sigma$ in 25 hours.

\begin{figure}[h]
\includegraphics[width=120mm]{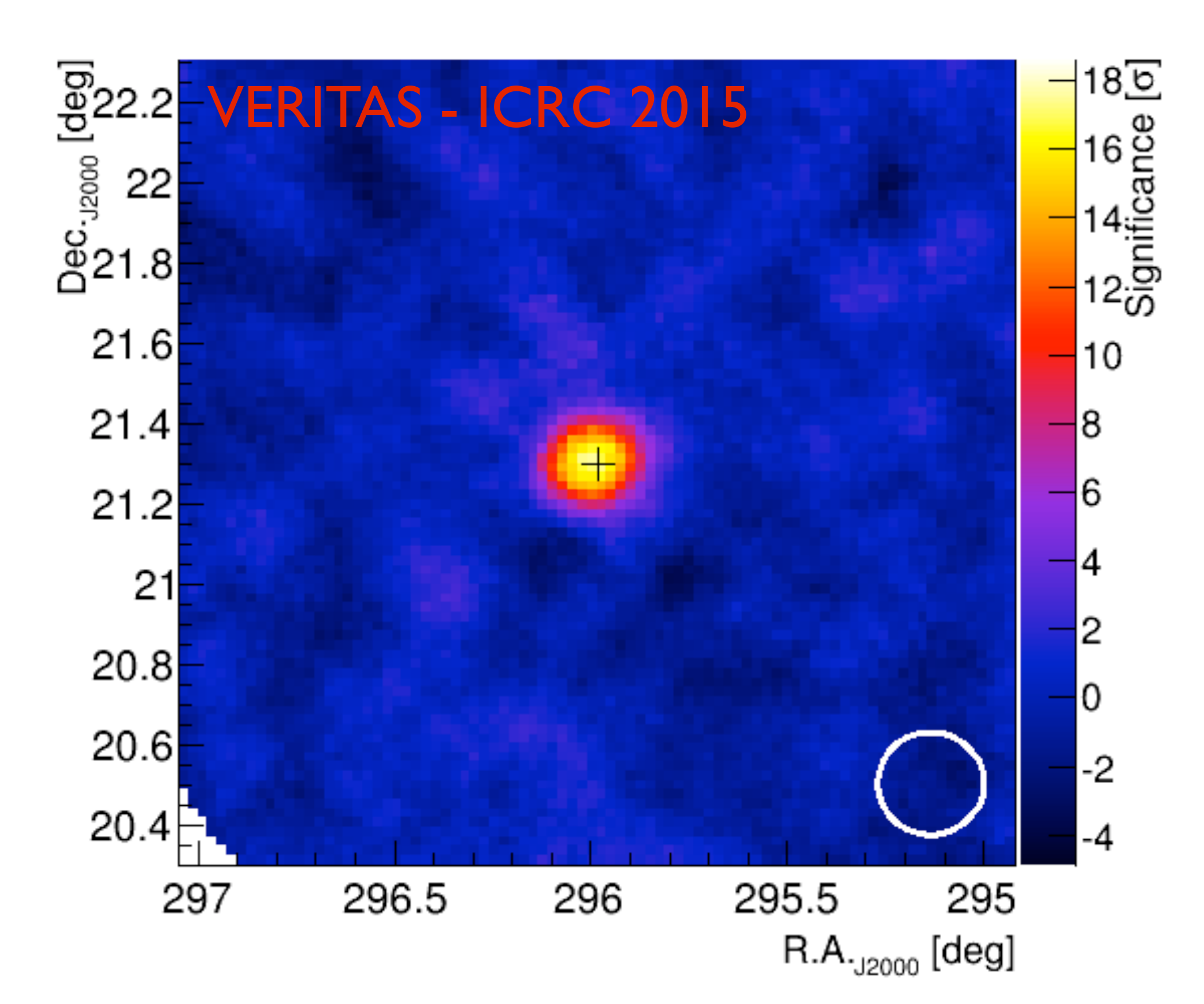}
\centering
\caption{Significance sky map of VERITAS observations of HESS J1943+213. The white circle represents the VERITAS angular resolution 68\% containment at 1 TeV; the cross shows the catalog position of the source.}
\label{sigmap}
\end{figure}

VERITAS observed HESS J1943+213 between May 27, 2014 (MJD 56804) and July 2, 2014 (MJD 56840) with 
22.5~hours of total live time after data affected by poor weather conditions are eliminated. Observations took place at elevations between 63$^{\circ}$ and 80$^{\circ}$, leading to 17.5$\sigma$ source detection above 200~GeV.

\begin{figure}[!h]
\includegraphics[width=100mm]{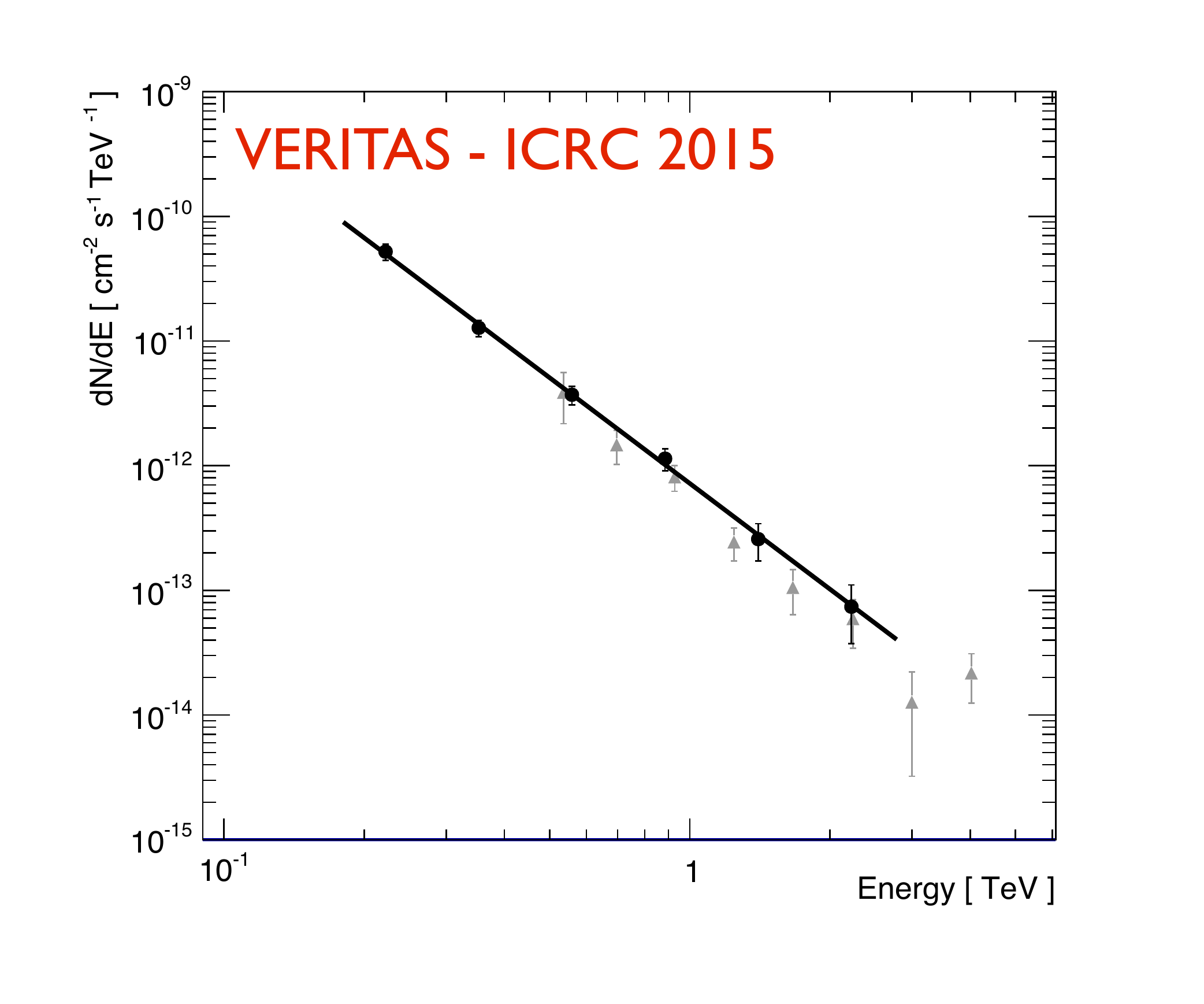}
\centering
\caption{Differential energy spectrum of HESS J1943+213. The spectrum with VERITAS (black points) is between 200~GeV and 2~TeV and is fit to a power-law function. The spectrum from the H.E.S.S. result \cite{hess} is also included for comparison (gray points).}
\label{spectrum}
\end{figure}

\begin{figure}[!h]
\includegraphics[width=100mm]{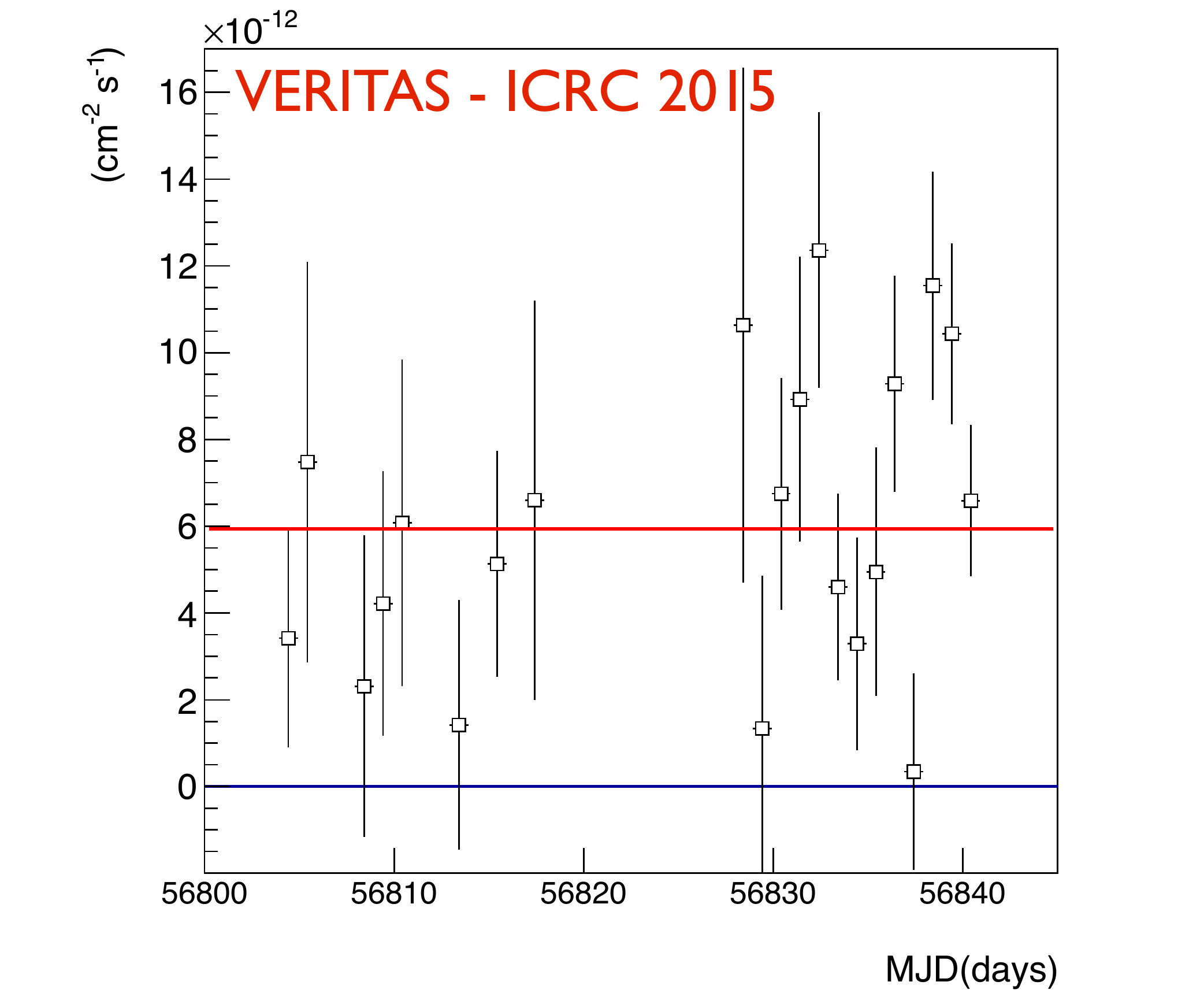}
\centering
\caption{Light curve of HESS J1943+213 with VERITAS, with flux measured above 200~GeV. The red line is a constant flux fit to the data.}
\label{lc}
\end{figure}

\section{Results}
Figure~\ref{sigmap} shows the significance sky map near HESS J1943+213, including the detected source. The source location is 
consistent with the catalog position of HESS J1943+213 (marked with a black cross).
The VERITAS differential energy spectrum of the source is shows in Figure~\ref{spectrum}. The high elevation observations lead to a lower energy threshold and a spectrum that extends down to 200~GeV, compared with 470~GeV from H.E.S.S. observations. The source spectrum is well fit by a power-law function with an index of $\Gamma$=2.82 $\pm$ 0.13 in the energy range 200~GeV--2~TeV. The VERITAS spectrum of HESS J1943+213 appears to be consistent with the spectrum from H.E.S.S. ($\Gamma$=3.1 $\pm$ 0.3), also depicted in Figure~\ref{spectrum}.

The flux of (1.30 $\pm$ 0.20) $\times$ 10$^{-12}$ cm$^{-2}$ s$^{-1}$ measured with VERITAS above 470 GeV is consistent with the flux of (1.25 $\pm$ 0.20) $\times$ 10$^{-12}$ cm$^{-2}$ s$^{-1}$ from the H.E.S.S. detection. As VERITAS is able to observe HESS J1943+213 at a much higher elevation than H.E.S.S., the detection rate of the source with VERITAS is $\sim$3.7$\sigma$/$\sqrt{\textrm{hour}}$, compared with $\sim$1.8$\sigma$/$\sqrt{\textrm{hour}}$ with H.E.S.S., allowing VERITAS to test for variability on a factor of four shorter timescales. 

Figure \ref{lc} illustrates the light curve of HESS J1943+214 observed with VERITAS, with fluxes measured above 200~GeV (showing only statistical uncertainties). A constant flux line is fit to the data (shown in red) with a $\chi^{2}$ / NDF = 31.34 / 20, corresponding to a p-value of 0.05. With systematic uncertainties in flux included, any hints of variability are not statistically significant.

\section{\emph{Swift}-XRT Observations and Results}

Soft X-ray observations of HESS J1943+213 were obtained with the \emph{Swift} XRT-instrument on 17 June, 19 June, and 21 June, 2014, contemporaneous with VERITAS observations. The data analysis for XRT observations was performed with the standard tools included in HEASoft package Version 6.15.1, specifically with the XRTDAS v3.0.0 tools, while spectral analysis and fitting was performed using Xspec v12.8.1g. 

XRT flux spectra were obtained by unfolding the counts spectra with instrument response functions 
included in CALDB 1.0.2 and by assuming an absorbed power-law functional form for the intrinsic 
spectrum.

\begin{figure}[!ht]
\includegraphics[width=100mm]{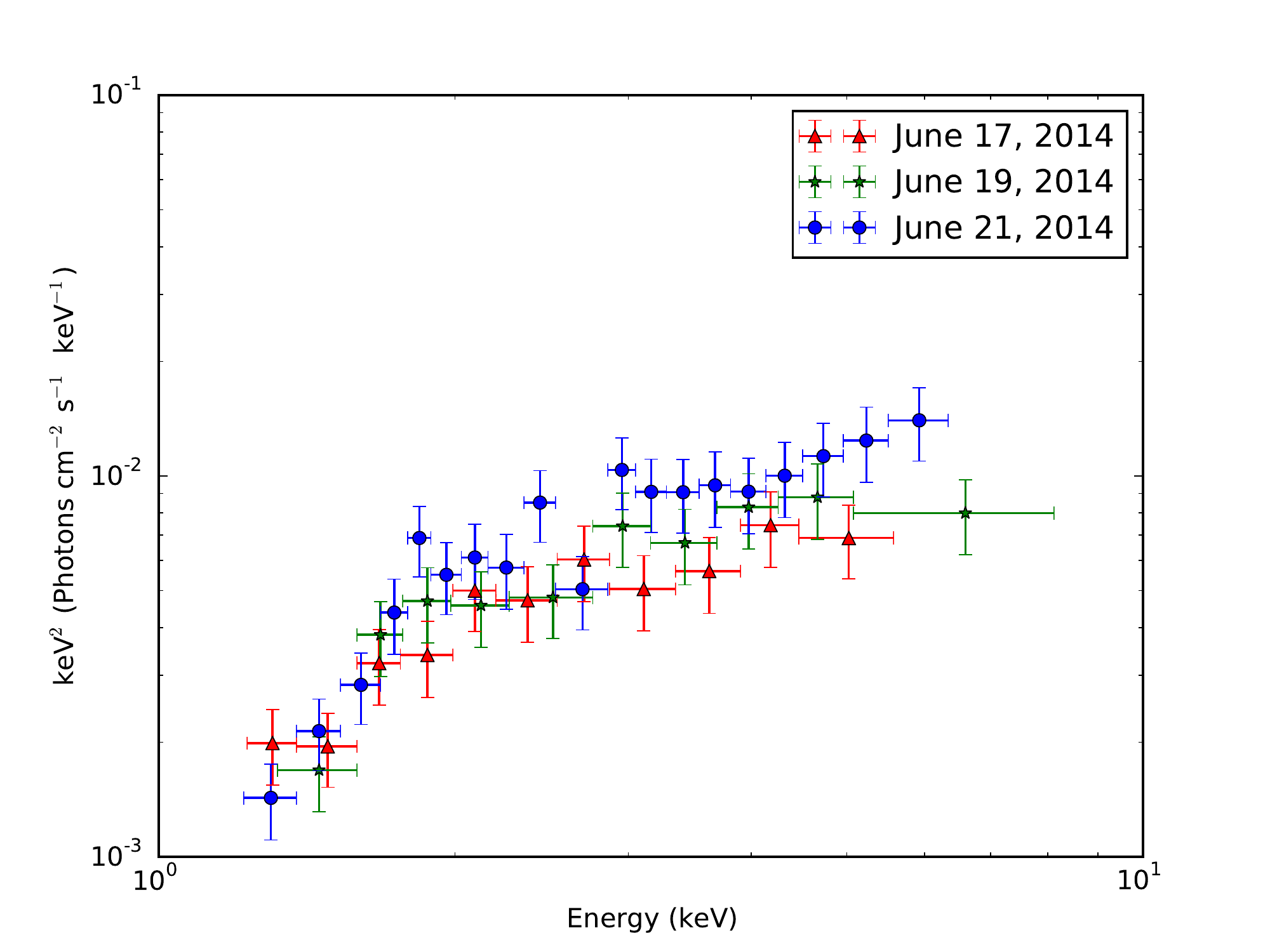}
\centering
\caption{Flux spectra of HESS J1943+213 with \emph{Swift} XRT for the three separate observations.}
\label{xrtspectra}
\end{figure}

The spectrum from each observation is shown in Figure \ref{xrtspectra} and the spectral fit information is included in Table~\ref{tab:xrt}. No significant spectral or flux variability is detected between observations. Moreover, the results are consistent with previous observations from \emph{Swift} XRT and other X-ray instruments in the 0.3--10 keV energy regime \cite{hess}. 

\begin{table}[!h]


\begin{tabular}{cccccc}
\hline
\bf{Date} & \bf{Observation ID} & \bf{Exposure} & \bf{Log10[Flux] (2--10 keV)} & \bf{Index} & \bf{$\chi$$^{2}$/NDF}\\
 & & (seconds) & Log10 (erg cm$^{-2}$ s$^{-1}$) & \\
\hline
2014-06-17 & 00033319001 & 967 & -10.74$\pm0.06$ & 1.76$\pm$0.43 & 3.29/8 \\
2014-06-19 & 00033319002 & 769 & -10.66$\pm0.05$  &  1.98$\pm$0.37 & 3.25/7 \\
2014-06-21 & 00033319003 & 1156 & -10.52$\pm0.04$  &  1.65$\pm$0.28 & 11.93/16 \\
\hline
\end{tabular}
\caption{Summary of \emph{Swift}--XRT observations.}
\label{tab:xrt}
\vspace{1 mm}
\end{table}

\section{Discussion and Outlook}

If HESS J1943+213 is a blazar, the agreement between VHE fluxes measured approximately five years apart with VERITAS and H.E.S.S.  and the lack of variability in the X-rays, including in the recent \emph{Swift} XRT result is surprising, but not unusual. Blazars are known to vary at all energies and at a wide range of timescales~\cite{boettcher}. The detection of variability at 1.5 GHz by ref.~\cite{gab2015} has weakened the lack of variability concern and is a strong piece of evidence in favor of the blazar scenario. 


VERITAS can probe timescales that are a factor of four shorter than those available to H.E.S.S and therefore possesses the best VHE dataset available for searches of flux and spectral variability from this source. Planned application of advanced analysis techniques, such as use of boosted decision trees for gamma-hadron separation will provide an even higher sensitivity and allow for an additional factor of two improvement in the minimum variability timescale that can be tested.

The VERITAS-measured spectrum of HESS J1943+213 is in good agreement with the H.E.S.S. spectrum and extends down to 200 GeV. The soft VHE spectral index constituted one of the key pieces of evidence against the PWN hypothesis. 

VERITAS will continue observations of HESS J1943+213 and will monitor the source for potential flux variability. In addition, the newly released PASS 8 Fermi-LAT data will allow for an improved detection and a spectrum of the source in GeV gamma rays, providing a significantly better handle on the gamma-ray peak of the HESS J1943+213 spectral energy distribution. Multi-wavelength studies of the source, including studies of its spectral energy distribution will be essential for definitively identifying the source class and investigating its properties.

\section{Acknowledgements}

This research is supported by grants from the U.S. Department of Energy Office of Science, the U.S. National Science Foundation and the Smithsonian Institution, and by NSERC in Canada. We acknowledge the excellent work of the technical support staff at the Fred Lawrence Whipple Observatory and at the collaborating institutions in the construction and operation of the instrument. The VERITAS Collaboration is grateful to Trevor Weekes for his seminal contributions and leadership in the field of VHE gamma-ray astrophysics, which made this study possible.

\end{document}